\documentclass[reprint,
nobibnotes,
 amsmath,amssymb,
 aps,
prl,
]{revtex4-2}
\usepackage[version=4]{mhchem} \usepackage{graphicx}% Include figure files
\usepackage{dcolumn}% Align table columns on decimal point
\usepackage{bm}% bold math
\usepackage{hyperref}% add hypertext capabilities
\usepackage{siunitx}
\usepackage{physics}

\begin{document}
\newcommand{\Mod}[1]{\ (\mathrm{mod}\ #1)}
\def\ddt{\frac{d}{dt}}
\def\kL{\ket{L}}
\def\kR{\ket{R}}
\def\kO{\ket{1}}
\def\kT{\ket{2}}
\def\kM{\ket{-}}
\def\kP{\ket{+}}
\def\bL{\bra{L}}
\def\bR{\bra{R}}
\def\bO{\bra{1}}
\def\bT{\bra{2}}
\def\bM{\bra{-}}
\def\bP{\bra{+}}
\def\ddx{\frac{d}{dx}}
\def\dddx{\frac{d^2}{dx^2}}
\def\ddt{\frac{d}{dt}}
\def\dddt{\frac{d^2}{dt^2}}
\def\omp{\omega'}
\def\rv{\textbf{r}}
\def\xv{\textbf{x}}
\def\yv{\textbf{y}}
\def\pv{\textbf{p}}
\def\fv{\textbf{f}}
\def\jv{\textbf{j}}
\def\lv{\textbf{l}}
\def\av{\textbf{a}}
\def\uv{\textbf{u}}
\def\bv{\textbf{b}}
\def\kv{\textbf{k}}
\def\wv{\textbf{w}}
\def\tt{ \theta}
\def\vv{\textbf{v}}
\def\Omv{\textbf{\Omega}}
\def\cross{\times}
\def\betav{\textbf{\beta}}
\def\epv{\mathbf{\epsilon}}
\def\vepv{\mathbf{\varepsilon}}
\def\Lang{\mathcal{L}}
\def\Ham{\mathcal{H}}
\def\Fv{\textbf{F}}
\def\omv{\bm{\omega}}
\def\Ev{\textbf{E}}
\def\Mv{\textbf{M}}
\def\Dv{\textbf{D}}
\def\Lv{\textbf{L}}
\def\Kv{\textbf{K}}
\def\Qv{\textbf{Q}}
\def\wv{\textbf{w}}
\def\Rv{\textbf{R}}
\def\Bv{\textbf{B}}
\def\Hv{\textbf{H}}
\def\Jv{\textbf{J}}
\def\Sv{\textbf{S}}
\def\grad{\textbf{\nabla}}
\def\ineps#1{\epsilon_{#1}}
\def\part{\partial}
\def\p{\phantom{o}}
\def\Av{\textbf{A}}
\def\la{\langle}
\def\ra{\rangle}
\def\cotan{\qopname\relax o{cotan}}
\def\ad{a^\dagger}
\def\st{\star}
\def\Bt{\mathcal{B}}
\def\eO{\epsilon_0}
\def\eps{\epsilon}
\DeclareSIUnit\rydberg{Ryd.}
\DeclareSIUnit \mob {\centi\meter\squared\per\volt\per\second}
\DeclareSIUnit \dens {\per\centi\meter\cubed}
\DeclareSIUnit \centimeter {\centi\meter}
\DeclareSIUnit \diff {\centimeter \squared \per \second}
\DeclareSIUnit \year {year}
\sisetup{ per-mode = fraction}
\def\D{\ \mathrm{d}}
\def\Ai{\ \mathrm{Ai}}
\def\Bi{\ \mathrm{Bi}}
\def\eo{\textbf{e_1}}
\def\et{\textbf{e_2}}
\def\fo{\textbf{f_1}}
\def\ft{\textbf{f_2}}
\def\rv{\textbf{r}}
\def\Rv{\textbf{R}}
\def\Kv{\textbf{K}}
\def\kv{\textbf{k}}
\def\qv{\textbf{q}}
\def\pv{\textbf{p}}
\def\jv{\textbf{j}}
\def\rhv{\hat{\rho}}
\def\grad{\nabla}
\def\part{\partial}
\def\sgo{\hat{\sigma}}
\def\Sv{\textbf{S}}
\def\norm{\hat{\textbf{n}}}
\newcommand{\parl}{{\mkern3mu\vphantom{\perp}\vrule depth 0pt\mkern2mu\vrule depth 0pt\mkern3mu}}
\def\tsub{\textsubscript}
\def\tsup{\textsuperscript}
\def\Id{\textbf{1}}
\def\sgx{\bm{\sigma}_x}
\def\sgy{\bm{\sigma}_y}
\def\sgz{\bm{\sigma}_z}
\def\sgv{\bm{\sigma}}
\def\rhov{\bm{\rho}}

\preprint{APS/123-QED}

\title{Enhanced Quasiparticle Relaxation in a Superconductor via the Proximity Effect}% Force line breaks with \\

\author{Kevin M. Ryan}
\affiliation{Northwestern University Department of Physics \& Astronomy, Evanston Illinois, 60208, USA}
\author{Venkat Chandrasekhar}%
\email{v-chandrasekhar@northwestern.edu }
\affiliation{Northwestern University Department of Physics \& Astronomy, Evanston Illinois, 60208, USA}

\

\date{\today}% It is always \today, today,

\begin{abstract}
Quasiparticle relaxation in pure superconductors is thought to be determined by the intrinsic inelastic scattering rate in the material. %, which at low temperatures is dominated by electron-electron scattering.
In certain applications, \textit{i.e.} superconducting qubits and circuits, excess quasiparticles exist at densities far beyond the thermal equilibrium level, potentially leading to dephasing and energy loss. % in such devices.
In order to engineer superconductors with shorter overall quasiparticle lifetimes, we consider the impact of a proximity layer on the transport of quasiparticles in a superconductor. We find that a normal metal layer can be used to significantly increase the relaxation rate of quasiparticles in a superconductor, as seen by a large reduction in the quasiparticle charge imbalance in a fully proximitized Cu/Al bilayer wire.
The mechanism for this effect may be useful for preventing quasiparticle poisoning of qubits using carefully chosen proximity bilayers consisting of clean superconductors and disordered normal metals.
\end{abstract}

\maketitle
% \section{Introduction}
Quasiparticle excitations are a persistent issue in the development of several superconducting technologies \cite{patel_phonon-mediated_2017,albrecht_transport_2017,karzig_quasiparticle_2021,svetogorov_quasiparticle_2022}, including superconducting qubits. Contributing to energy dissipation and dephasing, excess quasiparticles limit the lifetime of qubits and degrade their performance \cite{serniak_nonequilibrium_2019,glazman_bogoliubov_2021,aumentado_quasiparticle_2023}. Despite predictions that their density should be exponentially suppressed at the operating temperature of current superconducting quantum computers, many reports indicate populations of quasiparticles orders of magnitude higher than expected \cite{aumentado_nonequilibrium_2004,martinis_energy_2009,catelani_quasiparticle_2011,de_visser_number_2011,serniak_nonequilibrium_2019,glazman_bogoliubov_2021}. This vast discrepancy has lead to many proposals for the origin of these excess quasiparticles, ranging from infrared-light leaks and resonant coupling to microwave signals \cite{corcoles_protecting_2011,barends_minimizing_2011,liu_quasiparticle_2024}, stress induced phonon bursts \cite{anthony-petersen_stress-induced_2024}, cosmic radiation \cite{vepsalainen_impact_2020}, pair-breaking due to magnetic defects \cite{kumar_origin_2016,altoe_localization_2022} or by non-magnetic defects in the presence of intrinsic gap-anisotropy \cite{hirschfeld_electromagnetic_1989,tanatar_anisotropic_2022-1,zarea_effects_2023}. Often, these proposed mechanisms rely sensitively on materials properties of the supercondutor that are difficult (if not impossible) to engineer, or on extrinsic sources which cannot be mitigated via improved circuit design or avoided via quantum error correction \cite{wilen_correlated_2021}. As the origin of this ``quasiparticle poisoning'' continues to be debated, it seems unlikely that any one solution will prevent the generation of quasiparticles, and instead, new focus is being placed on engineering their dynamic behavior and relaxation. 

The study of how quasiparticles diffuse and relax within superconductors is an old one, dating back to original works by Clark and Tinkham \cite{clarke_experimental_1972,tinkham_theory_1972,chi_quasiparticle_1979,chi_addendum_1980} and others \cite{lemberger_charge-imbalance_1981,stuivinga_non-equilibrium_1981} exploring the relaxation rate of quasiparticles in superconductors among other non-equilibrium effects \cite{pals_non-equilibrium_1982}. In many of these studies, the superconductors are deliberately driven out of equilibrium by an injected current from a normal metal to either enhance  \cite{gray_enhancement_1978} or suppress \cite{fuchs_energy_1977} their superconductivity. As a net charge is injected, this also leads to an imbalance between electron- and hole-like quasiparticle populations that shifts the chemical potential for quasiparticles vs. Cooper-pairs \cite{pethick_relaxation_1979,pals_non-equilibrium_1982,tinkham_introduction_2004,cadden-zimansky_charge_2007}, permitting simultaneous control over several degrees of freedom of their non-equilibrium distribution. Only relatively recently have these types of experiments been performed at \SI{}{\milli\kelvin} temperatures \cite{cadden-zimansky_charge_2007,hubler_charge_2010,wolf_spin_2013,kuzmanovic_evidence_2020}, but these well established techniques for generation and detection of quasiparticles have not been implemented to study quasiparticle poisoning in superconducting qubits.  

Instead, much work currently focuses on a variety of quasiparticle trapping schemes intended to divert quasiparticles within qubits using normal metals. Such experiments typically focus on extracting quasiparticles into discrete reservoirs which are only weakly attached to the superconductor  \cite{pekola_trapping_2000,lang_banishing_2003,riwar_normal-metal_2016}, with their efficacy determined via qubit operation. Several recent works attempting to mitigate other issues in superconducting qubits, such as surface-oxide induced dielectric loss, have also begun to successfully employ normal metal encapsulation of the entire surface of the device \cite{chang_eliminating_2024, bal_systematic_2024-1,de_ory_low_2024}. 
In both cases, little information can be extracted about the quasiparticles themselves. This is especially relevant for encapsulated qubits, where the proximity effect present along the extended, highly transparent interface raises questions about how quasiparticles may behave dynamically in the presence of a spatially non-uniform superconducting gap. While it has already been proposed that gap suppression induced by the inverse proximity effect can have either a positive or negative effect on transmon qubit performance depending on the sample geometry \cite{riwar_dissipation_2018,hosseinkhani_optimal_2017,riwar_efficient_2019}, relatively little work has directly examined the relaxation of non-equilibrium quasiparticles directly within such a bilayer. To explore this issue, we present in this Letter a comparative study of charge imbalance in a simple Al device and one consisting of an Al/Cu (20nm/3nm) proximity effect bilayer. Using tunneling methods derived from the foundational studies in non-equilibrium superconductivity, our results demonstrate the potential for engineering enhanced quasiparticle relaxation in a proximity effect system using a method compatible with qubit fabrication.

Charge and energy imbalance, being the first observed signatures of non-equilibrium superconductivity, can provide direct access to inelastic scattering rates/lifetimes for quasiparticles in a given system. In most studies, charge imbalance is generated via injection of quasiparticles of charge $q_k$ (being always less than or equal to the ordinary electron/hole charge) through normal-metal/insulator/superconductor (NIS) tunnel junctions. In Clark's original experiment \cite{clarke_experimental_1972}, NIS and SIS (Josephson) junctions were co-located on the same superconducting wire, and used to drive and measure a steady-state, non-equilibrium quasiparticle distribution $f_k(\epsilon_k)$ \cite{pethick_relaxation_1979,tinkham_introduction_2004}. In this procedure, when injecting a current of predominantly electron or hole-like Bogoliubov quasiparticles across the NIS junction, one can calculate the accrued charge imbalance $Q^*$ by integrating the total charge of all quasiparticles 
\begin{align}\label{eq:Qdef}
    Q^* \equiv \int_{-\infty}^{\infty} q_k  f_k(\epsilon_k) d\epsilon_k.
\end{align}
This total charge is compensated by a change in the chemical potential for quasiparticles in the superconductor \cite{pethick_relaxation_1979,tinkham_introduction_2004}
\begin{align}
    \mu_{Q^*} = \frac{Q^*}{2eN(0) },
    \label{eq:muq}
\end{align}
where $N(0)$ is the normal-metal density of states at the Fermi energy, which can be independently detected and compared to the chemical potential for Cooper pairs. This is often done by observing the current across a second NIS junction, which is given by $I = G^N \times \mu_{Q^*}$, where $G^N$ is the normal state conductance of the junction.  From Eq. \ref{eq:Qdef}, this will contain contributions from all energies, being non-zero only in the presence of charge imbalance.

In 2010 Hübler \textit{et al.} \cite{hubler_charge_2010} directly measured the real space relaxation of quasiparticle charge imbalance at \SI{}{mK} temperatures relevant for superconducting quantum applications. Using pairs of NIS tunnel junctions, one biased at a finite voltage and the other held at zero bias, they demonstrated that a non-local current due to charge imbalance alone could be detected away from the injector junction with sufficient amplitude to determine the energy dependence of the charge imbalance relaxation length in Al at low temperatures. To summarize the process: first, a current $I_{inj}(V_{inj})$ is injected at a given bias to generate a charge imbalance $Q^*$ by tunneling of electrons or holes into the superconductor which then relax at a rate of $1/\tau_{Q^*}$. For large injector currents, a large charge imbalance $Q^*$ will be established resulting in a finite $\mu_{Q^*}$ according to Eq. \ref{eq:muq}. $\mu_{Q^*}$ decays exponentially over the charge imbalance length
\begin{equation}
    \lambda_{Q^*}^2 = D(\epsilon_k) \tau_{Q^*} =  \frac{\epsilon_k}{\sqrt{\epsilon_k^2 + \Delta
    ^2}} \times D_N\tau_{Q^*},\label{eq:DiffusionCoeff}
\end{equation}
where $D(\epsilon_k)$ is the diffusion constant for quasiparticles and $D_N$ is diffusion constant for electrons in the normal state \cite{pethick_relaxation_1979,ullom_energy-dependent_1998}. Finally, one observes the steady-state quasiparticle current $I_{det} = G^N_{det} \times \mu_{Q^*}$ using a second NIS junction at zero-bias a distance $L$ from the injector.  By measuring at zero detector bias, this current stems entirely from the chemical potential within the superconductor due to charge imbalance alone, and by making multiple measurements at varying $L$, the exponential relaxation can be traced out. More detailed derivations of the quasiparticle conductance between junctions have been presented extensively elsewhere in the literature \cite{pethick_relaxation_1979,pals_non-equilibrium_1982,tinkham_introduction_2004,hubler_charge_2010,kolenda_nonlocal_2013}. 

 \begin{figure}[b]
    \centering
    \includegraphics[width=\linewidth]{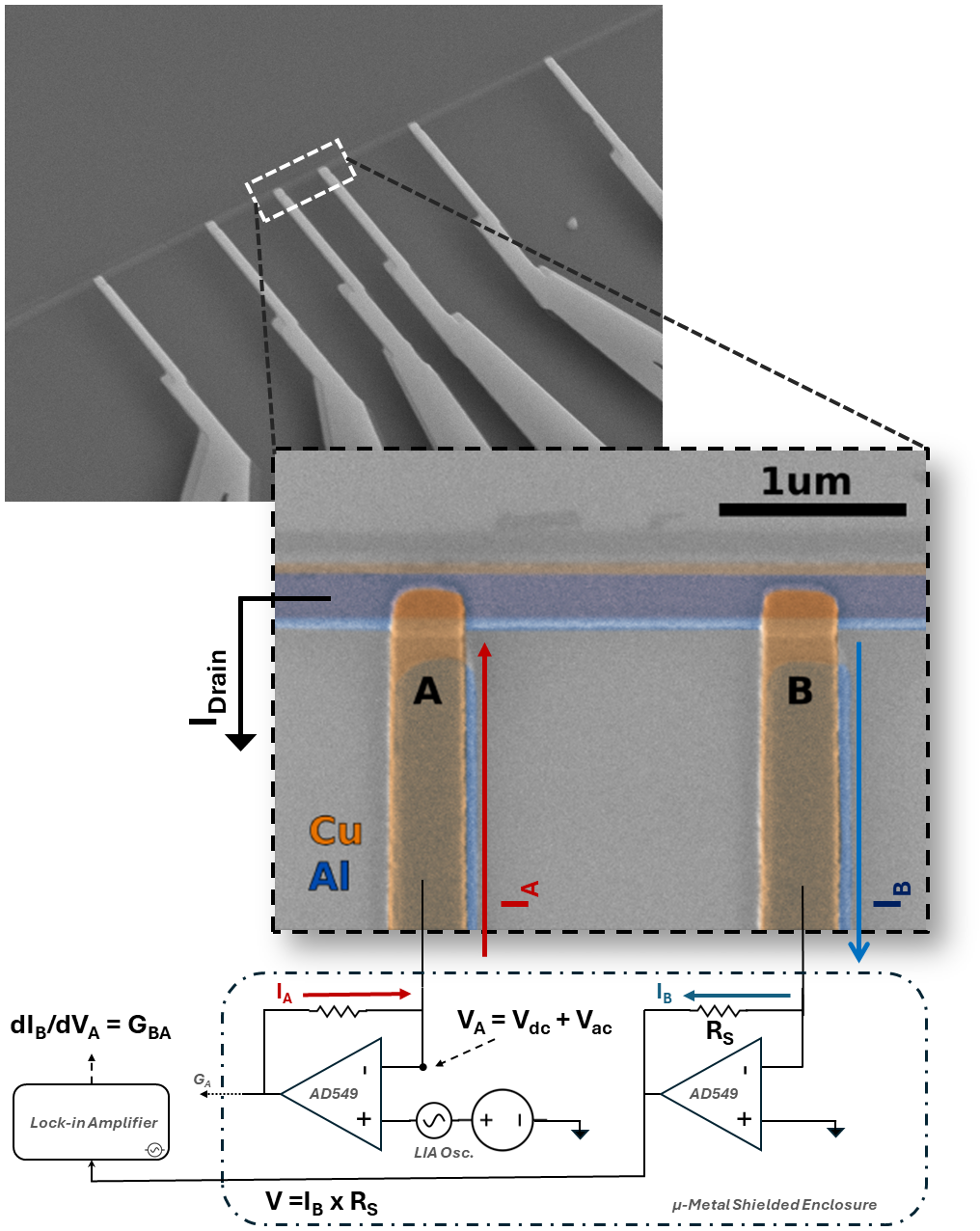}
    \caption{Top: False-color SEM micrograph of the inner-most pair of leads on the bilayer device, and a circuit diagram of the voltage biased current pre-amplifiers used for conductance measurements.}%, and the potential for supression of the gap towards the interface of the NS bilayer to cause preferential quasiparticle flow there.}
    \label{fig:Schematics}
\end{figure}
In order to compare the charge imbalance relaxation both with and without the influence of the inverse proximity effect, a pair of devices consisting of 6 junctions spaced between 2 and \SI{6}{\micro\meter} apart with a maximum separation of \SI{20}{\micro\meter} were fabricated using a conventional PMGI/PMMA bilayer e-beam lithography process on 1um \ce{SiO_x} substrates. A false color SEM micrograph of the Al/Cu bilayer device is shown as part of Fig. \ref{fig:Schematics}. All metal depositions and oxidation steps were performed without breaking vacuum, ensuring a transparent interface between the Cu and Al. Only the desired Al or Al/Cu bilayer wire spans the full width of the junction region. This is done by by using a very minimal PMGI undercut so that metalization layers are deposited on the resist sidewall and removed during liftoff. Cu was chosen as a normal metal and tunnel injector over noble metals such as Au used in recent encapsulated qubits due to the potential for forming insulating Al intermetallic interfaces. We find that the Al/Cu interface is remarkably stable, giving transparent NS contacts even after junction aging. Liftoff was performed with alternating baths of hot acetone and an NMP based remover, followed by ultrasonication in acetone and isopropanol to remove the metalized sidewall. The substrates were then cleaved to size, imaged via SEM, and wirebonded before being cooled to \SI{77}{\kelvin} within 48 hours of fabrication.  More details of the fabrication process are given in the End Matter.

Measurements were performed using a $^3$He/$^4$He dilution refrigerator with a base temperature of approximately \SI{19}{\milli\kelvin}. Each measurement line has ~\SI{20}{\ohm} of series lead resistance at base temperature, and is filtered with an \SI{800}{\kilo\hertz} low-pass $\pi$-filter. All pre-amplifiers were battery powered and housed in a $\mu$-metal enclosure immediately beside the dilution refrigerator and connected to it via a $\mu$-metal shielded cable. Both grounding and vibration isolation were carefully considered and were essential to obtaining low electronic temperatures given the large number of connections for each device.

Both single and dual-junction conductance measurements were performed using custom voltage biased transimpedance amplifiers based on AD549 electrometer op-amps with an input impedance of \SI{E13}{\ohm} and a gain of \SI{E7}{V/A}. To improve the signal to noise, a combined ac+dc bias was produced via a lock-in amplifier and function generator, with ac excitation of 1 to \SI{5}{\micro\volt} depending on the full scale sensitivity for a given measurement. A schematic of the dual-conductance measurement used is shown in Fig. \ref{fig:Schematics}. In short, independent dc-biases are applied to a pair of junctions A/B with ac-excitation at frequency $f_A$ and $f_B$ respectively.  As the junction A is closer to the ground point than junction B, any change in current measured at junction B in response to bias changes at junction A is a nonlocal response. Measurements of the junction conductances $G_A$ and $G_B$ were also performed to characterize the local behavior of the junctions. Note that, due to input voltage offsets, small bias corrections of 10 to \SI{20}{\micro\volt} have been made to ensure the symmetry of $G_A$ and $G_B$. The resistance across the Al or Al/Cu wire itself was obtained via a 4-terminal resistance measurement with an excitation of \SI{100}{nA}.

\begin{figure}[t]
    \centering
    \includegraphics[width=1.0\linewidth]{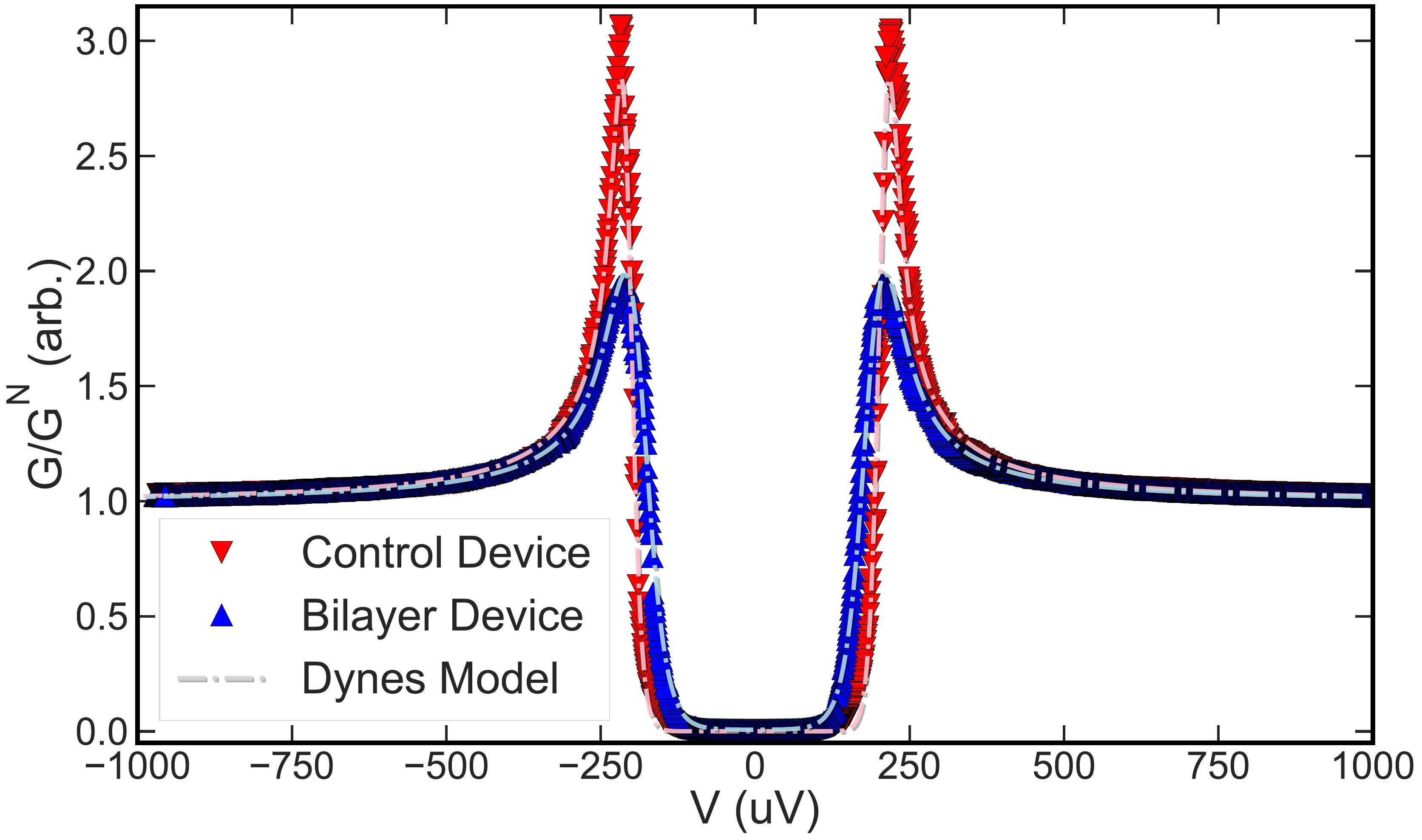}
    \caption{Normalized local conductance and Dynes-type fits for two junctions, one from the control and and one from the bilayer device, taken with identical measurement configurations at a bath temperature of \SI{20.4}{mK}.  \textit{Fit parameters---} Control: $T_e = \SI{101}{mK}$, $G^N=\SI{202}{\micro S}$, $\Delta=\SI{207}{\micro eV}$, and $\Gamma < \SI{E-5}\Delta$. Bilayer: $T_e = \SI{196}{mK}$, $G^N=\SI{152}{\micro S}$, $\Delta=\SI{189}{\micro eV}$, and $\Gamma < \SI{0.6}{\%}\Delta$.}
    \label{fig:ProxDOS}
\end{figure}

To characterize the quality of the NIS junctions, we fit their local conductance $G=dI/dV$ as a function of junction bias $V$ to the conventional theory with a broadened density of states characterized by Dyne's parameter $\Gamma$, with $G_N$, $\Delta$, $\Gamma$ and the effective electron temperature $T_e$ as fitting parameters \cite{dynes_direct_1978,meservey_spin-polarized_1994}. Fig. \ref{fig:ProxDOS} shows representative fits for junctions on both the control and bilayer devices.  For the control device, the Dynes model fit yields an effective electronic temperature of $T_e \approx\SI{100}{mK}$ with a value of $\Gamma<10^{-5}\Delta$.  If we fix $T_e$ to be the measured mixing chamber temperature of 20 mK, $\Gamma\sim 0.03 \Delta$, but the two fits are essentially indistinguishable on the scale of the plots in Fig. \ref{fig:ProxDOS}.  For both fits, the value of $\Delta = 207$ $\mu$eV obtained is the same, and this is the parameter that we use in our analysis below.  For the bilayer, the unconstrained fit gives $T_e =\SI{196}{mK}$.  This is most likely due to spatial inhomogeneity in the gap $\Delta$ across the junction caused by the intentional offset between the Cu and Al layers, rather than actual thermal broadening. Thus, tunneling in both devices is in good agreement with the predicted BCS density of states for the control and a with a proximity effect in the bilayer consistent with a high transparency interface.

\begin{figure}[]
    \includegraphics[width=0.48\textwidth]{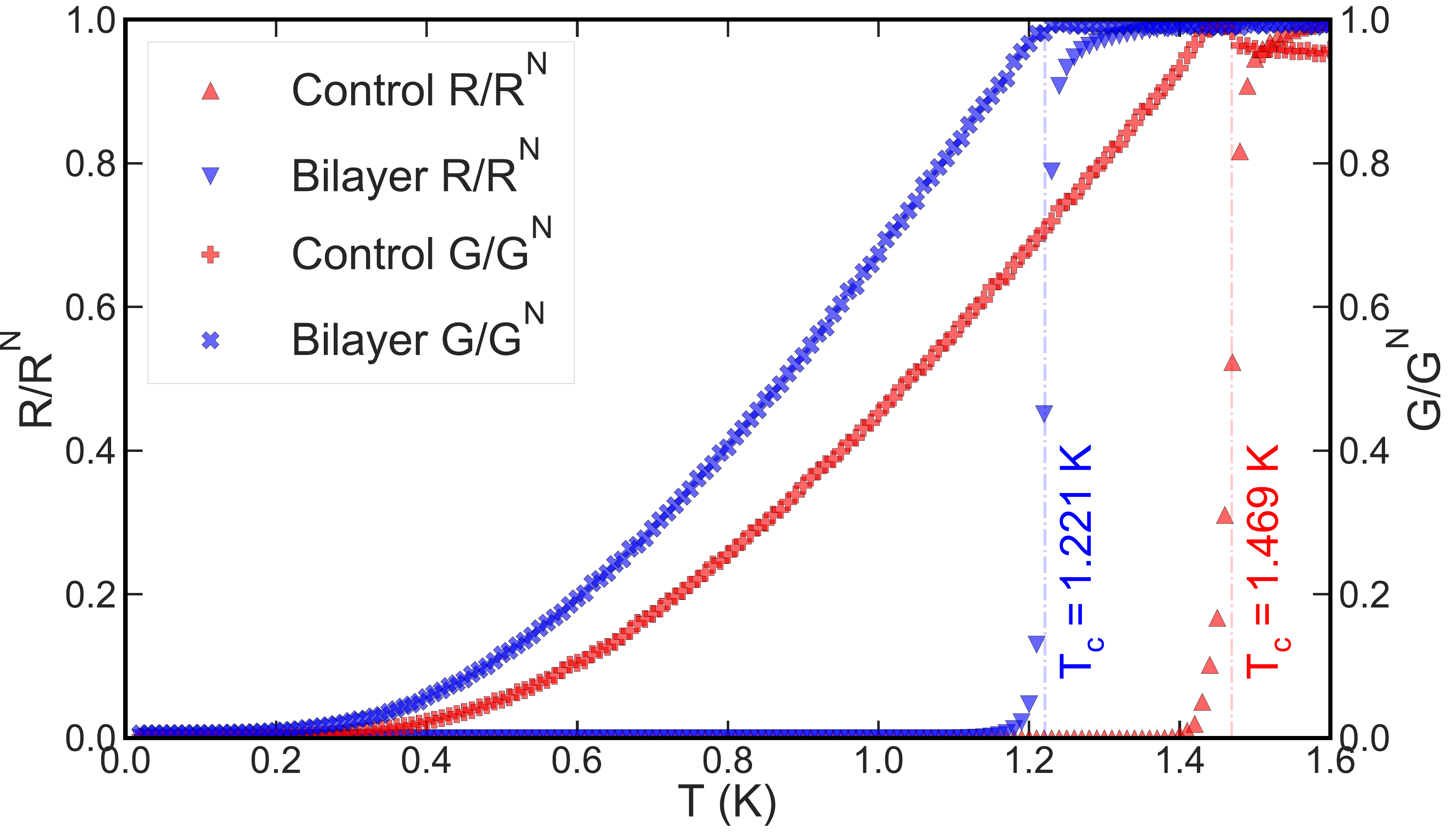}
    \caption{Resistance of the superconducting/bilayer wire, and conductance of their corresponding junctions normalized to their maxima below 1.6K. }
    \label{fig:Tc}
\end{figure}

Along with suppression of the gap, the inverse proximity effect causes a corresponding decrease in $T_c$, shown in Fig. \ref{fig:Tc}. The inclusion of the Cu layer causes a reduction in $T_c$ from \SI{1.47}{K} to \SI{1.22}{K} (taking $T_c$ to be the value where $R/R^N = 0.5$), a suppression of $\approx17\%$. Assuming a transparent NS interface, calculations based on the Usadel equation predict a suppression of $13\%$ for an ideal \SI{3}{\nano\meter}/\SI{20}{\nano\meter} NS bilayer\cite{chandrasekhar_proximity-coupled_2008}. The difference might be due to the slight offset between the layers that causes an enhancement of the inverse proximity effect, in agreement with the tunneling density of states which suggests a spatial variation of the gap.

% \subsection{Charge Imbalance in Pure Al Device}
Next we consider the signatures of charge imbalance in dual junction experiments that measure the nonlocal conductance $G_{BA}=dI_B/dV_A$ for both devices. As both the injected and extracted quasiparticle currents scale in magnitude with the intrinsic conductance of the junction, we must account for differences between chosen junctions in addition to their separation to fully describe the charge imbalance effect. To do so, we employ the model of Hübler \textit{et al.} \cite{hubler_charge_2010}, which provides a functional form for the conductance between junctions separated by a distance $L$:
\begin{equation}\label{eq:HublerModel}
    G_{BA} = g^*(eV_A) G_B^N G_A^N \frac{\rho_N \lambda_{Q^*}}{2A} e^{-L/\lambda_{Q^*}},
\end{equation}
or equivalently: 
\begin{equation}\label{eq:HublerModelNorm}
    \frac{2l}{R_N} \frac{G_{BA}}{G_A^N G_B^N} = g^*(eV_A) \lambda_{Q^*} e^{-L/\lambda_{Q^*}},
\end{equation}
where $A$ and $l$ are the cross sectional area and length of the central wire, $G_A^N$ and $G_B^N$ are the normal state local conductances of junctions A and B, $L$ is the distance between the junctions, $\rho_N$ and $R_N$ are the normal state resistivity  and resistance of the wire and $g^*$ is a dimensionless shape function on the order of unity associated with the BCS density of states and thermal broadening, being approximately zero for $eV_A<\Delta$ when no quasiparticle current is injected \cite{hubler_charge_2010}. 

\begin{figure}[]
\centering
\includegraphics[width=0.48\textwidth]{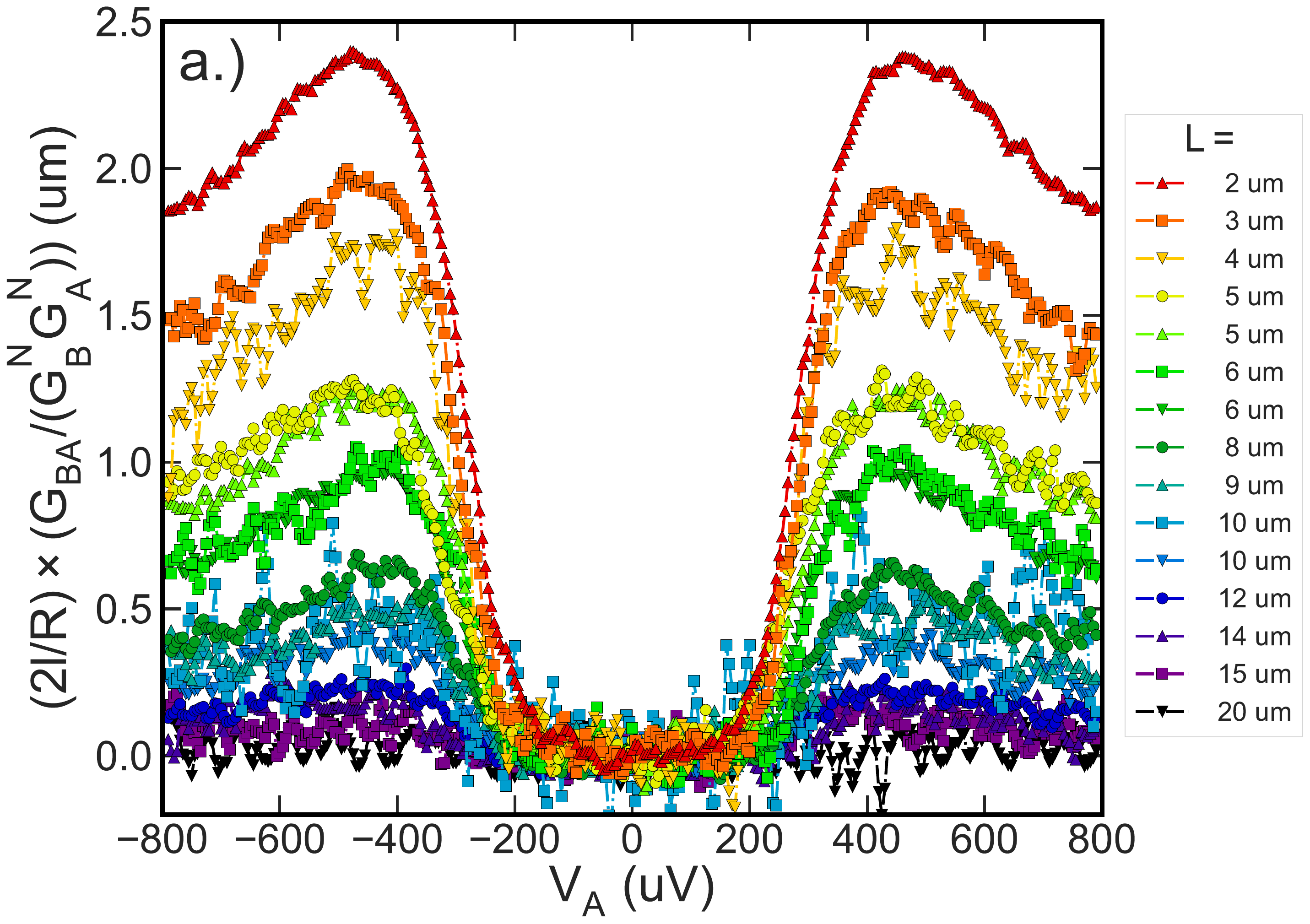}\\
\includegraphics[width=0.48\textwidth]{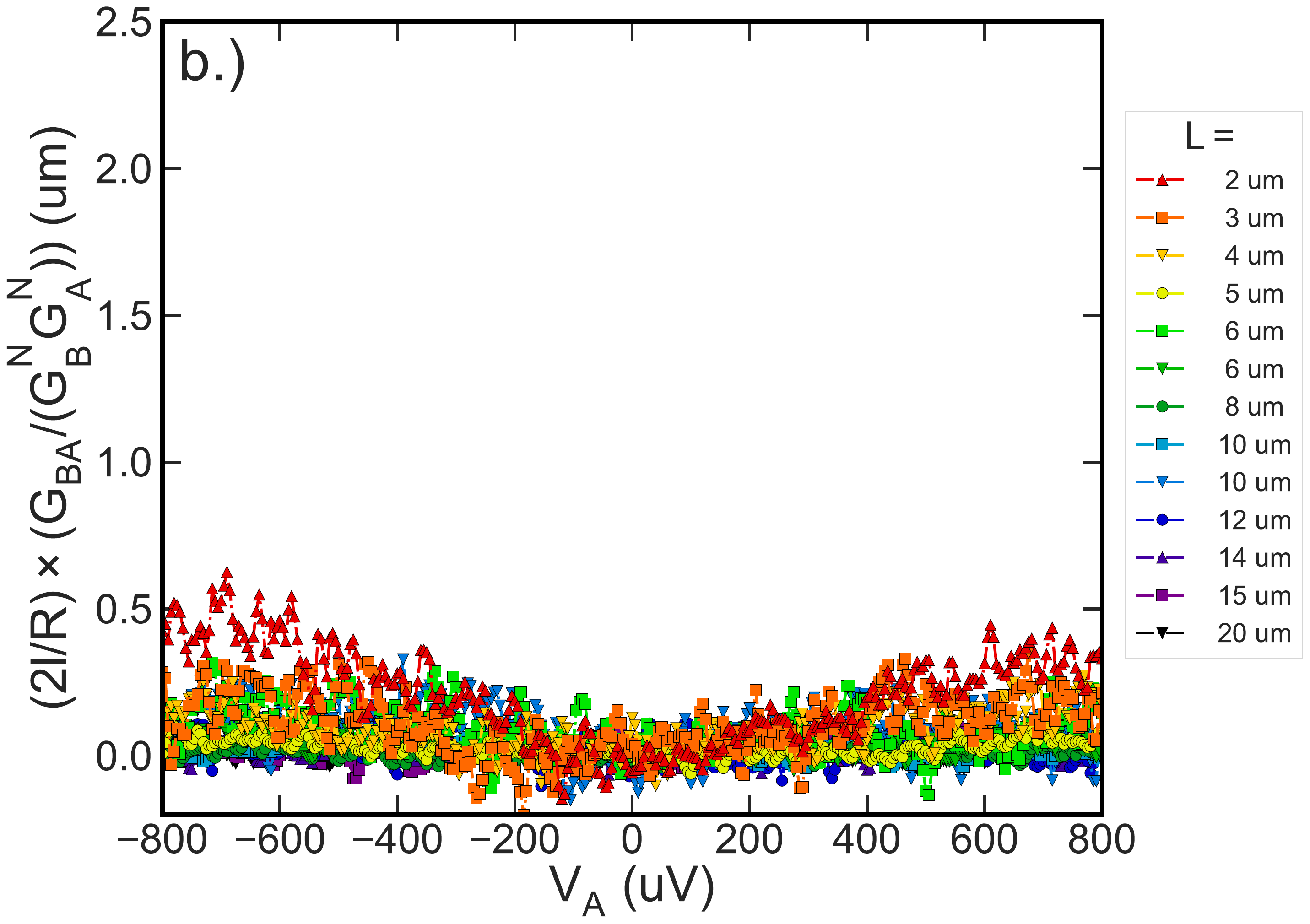}
\caption{Quasiparticle conductance normalized by Eq. \ref{eq:HublerModelNorm} for different values of the length $L$ between the injector and detector junctions, measured at \SI{20.7}{mK} for both a.) the control and b.) bilayer device.} 
\label{fig:NormCondComp}
\end{figure}

The normalization performed in Eq. \ref{eq:HublerModelNorm} is convenient as it accounts for differences in the resistance of the wires, and eliminates the cross sectional area $A$ which is harder to determine than the wire length of $l=\SI{22}{\micro\meter}$ for both devices. These normalized data are shown in Figs. \ref{fig:NormCondComp}a and \ref{fig:NormCondComp}b, plotted on the same scale to emphasize the difference between the devices. As expected for charge imbalance, a signal only appears for $V_A>\Delta$ and the data are generally symmetric with respect to $V_A$ \cite{pethick_relaxation_1979,tinkham_introduction_2004,hubler_charge_2010}. In the control device we can observe a nonlocal conductance over $L\leq \SI{15}{\micro\meter}$.  In the bilayer device, the signal is drastically reduced, being clearly discernible only for $L<\SI{4}{\micro\meter}$.  A similar suppression was observed in a second device measured at \SI{300}{\milli \kelvin}.

\begin{figure}[t]
\centering
\label{fig:ControlLength}\includegraphics[width=0.48\textwidth]{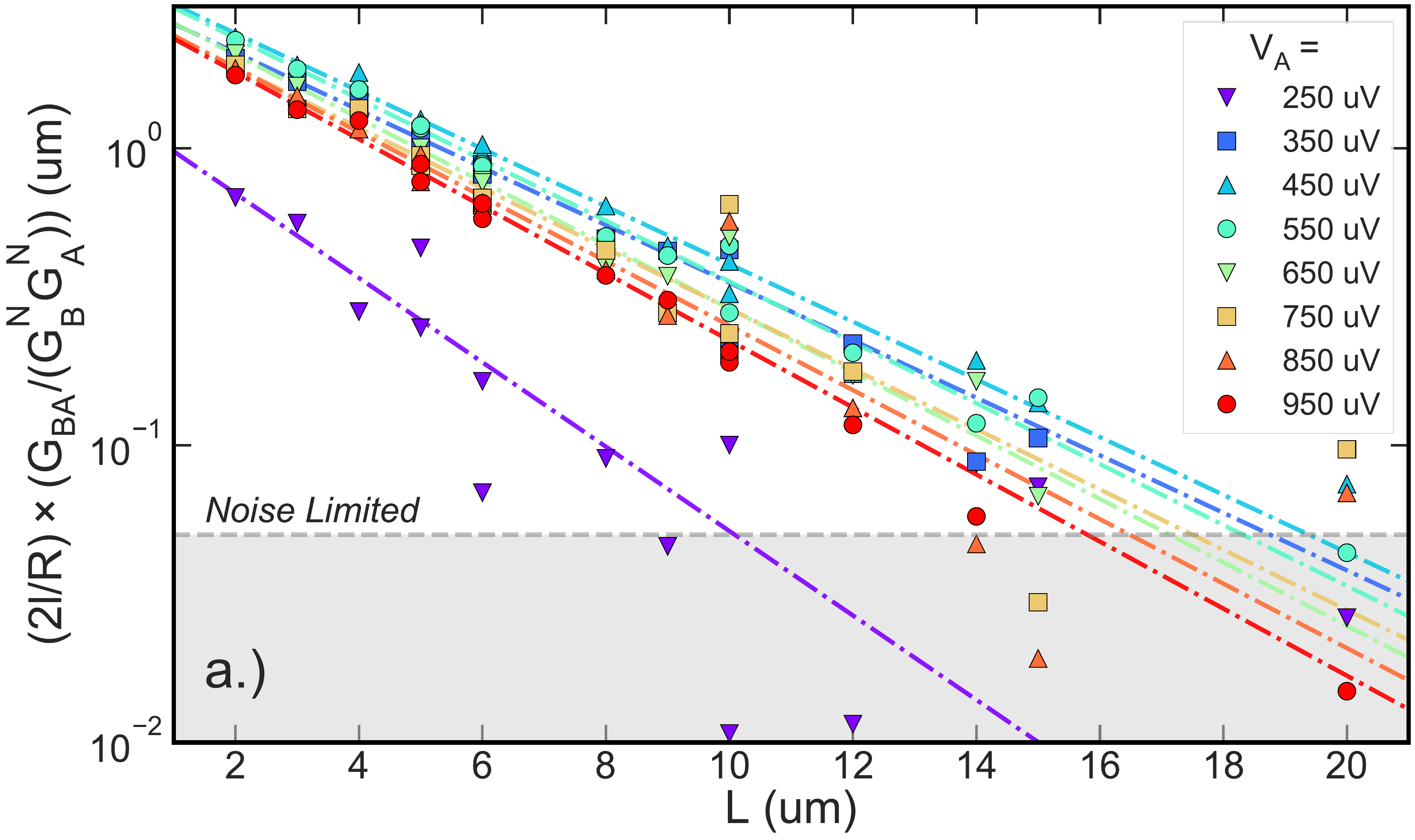}\\
\label{fig:BilayerLength}\includegraphics[width=0.48\textwidth]{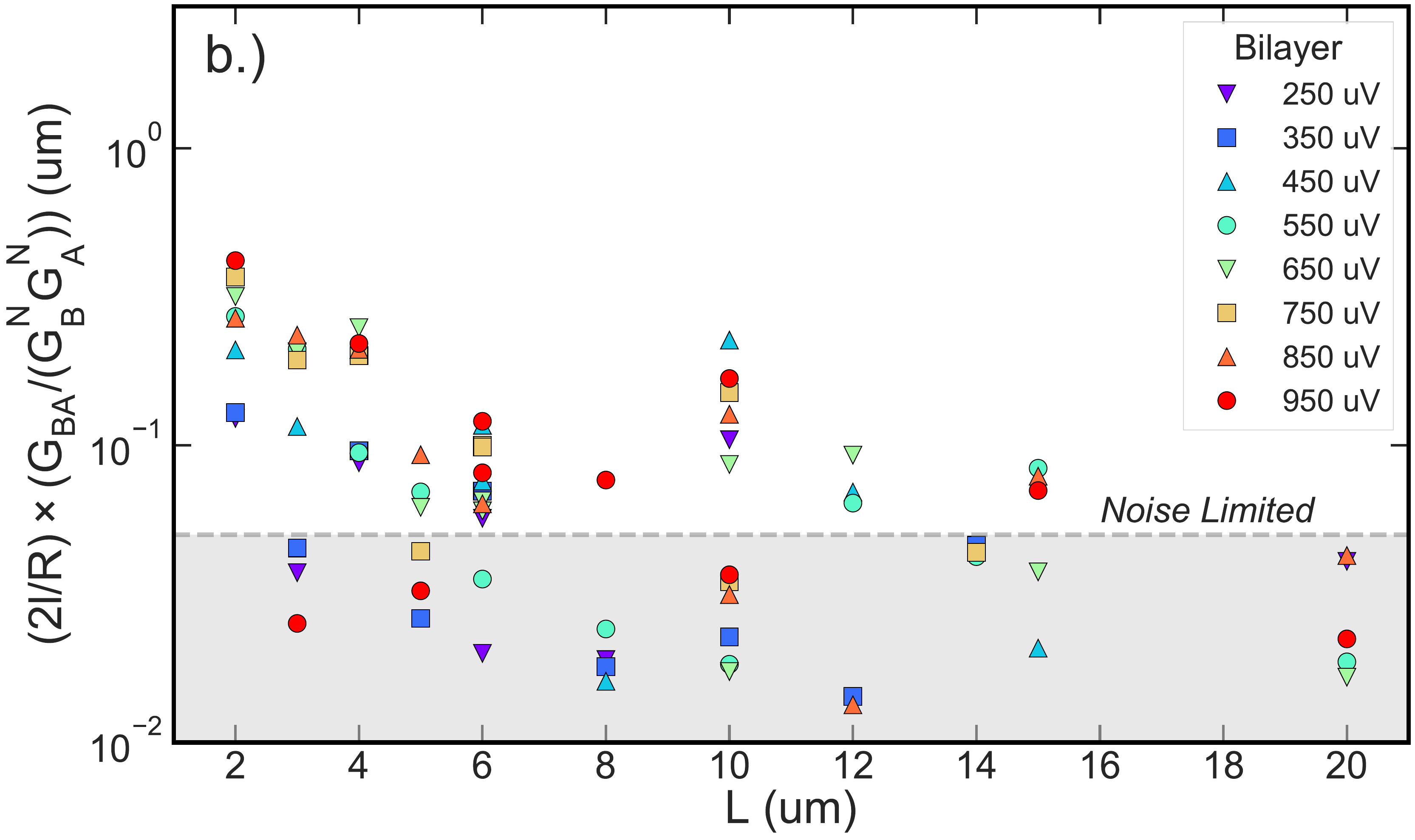}
\caption{ Normalized nonlocal conductance extracted from the data in Fig. \ref{fig:NormCondComp} at specific values of $V_A$ plotted logarithmically as a function of junction separation $L$ for a.) the control and b.) bilayer device. The dashed lines for the control device are fits to Eq. \ref{eq:HublerModelNorm} for values $eV_A \gtrapprox \Delta$. The grey regions in both figures defines a noise floor for the measurement of $\approx\SI{0.05}{\micro \meter}$.} 
\label{fig:LengthComp}
\end{figure}

To extract the length dependence of the charge imbalance signal, we can perform cuts of the data at specific injector voltages and plot them logarithmically as a function of $L$ as shown in Figs. \ref{fig:LengthComp}a. and \ref{fig:LengthComp}b. Plotted in this manner, the exponential relaxation over an energy dependent length $\lambda_{Q^*}$ is clearly seen in the Al control device in agreement with previous reports \cite{hubler_charge_2010}.  The signal for the bilayer device is reduced by over an order of magnitude below the control device, and falls below our noise threshold for $L>4$ $\mu$m.  

\begin{figure}[]
    \centering
    \includegraphics[width=1\linewidth]{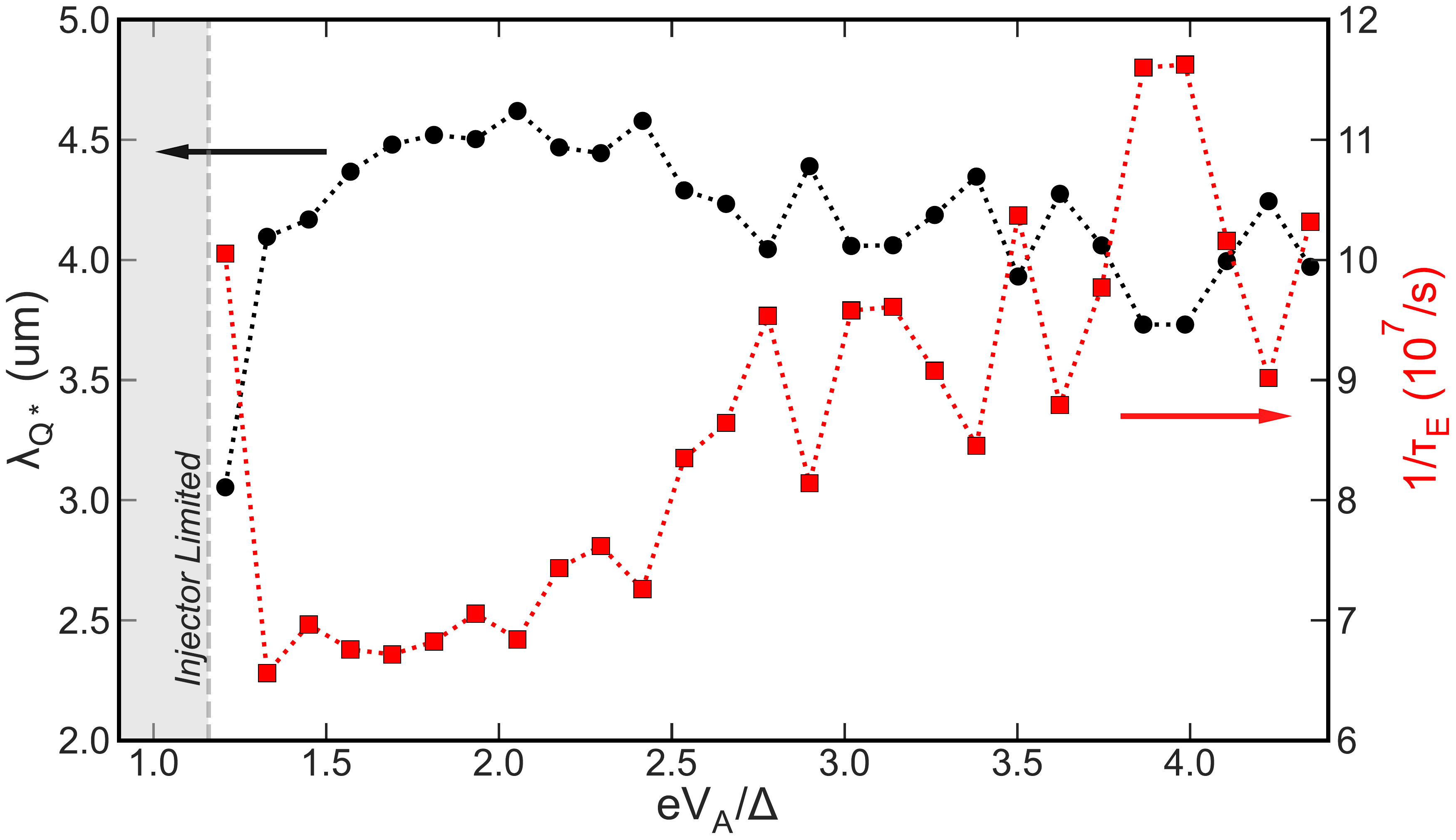}
    \caption{Charge imbalance length $\lambda_{Q^*}$ (black circles) for the control device as a function of injector bias $V_A$  obtained by fitting the data of Fig. \ref{fig:LengthComp}a  to Eq. \ref{eq:HublerModelNorm}.  The red squares show the corresponding $\tau_E$ values determined from $\lambda_{Q^*}$ using Eq. \ref{eq:taurelation} and Eq. \ref{eq:DiffusionCoeff}, assuming an excitation energy of $eV_A$. The grey region indicates energies where insufficient current $I_A$ is injected to measure $\lambda_{Q^*}$.}
    \label{fig:LambdaQStar}
\end{figure}

For the control device, we can fit the data in Fig. \ref{fig:LengthComp}a to Eq. \ref{eq:HublerModelNorm} to obtain an estimate of $\lambda_{Q^*}$. This is shown in Fig. \ref{fig:LambdaQStar} for different values of $V_A$. We obtain charge imbalance lengths between 4 and \SI{4.5}{\micro\meter} for the control device.  For the bilayer device, due to the greatly reduced charge imbalance signal, it is not possible to reliably extract $\lambda_{Q^*}$ with the same method. However, assuming the exponential dependence in Eq. \ref{eq:HublerModelNorm}, comparing the magnitudes of the signals of the control and bilayer devices at $L=\SI{2}{\micro\meter}$, and knowing the value of $\lambda_{Q^*}$ for the control device, we can estimate that $\lambda_{Q^*}$ in the bilayer device is approximately \SI{1.3}{\micro \meter} or less. 

% % \section{Discussion}

The drastic reduction of the charge imbalance signal in the bilayer device clearly demonstrates that the introduction of a simple proximitized Cu layer to an Al superconducting wire can cause dramatic enhancement of the relaxation of quasiparticles and their charge imbalance. At the same time, the bilayer remains robustly superconducting at base temperature. We can understand this effect by considering the role of both layers in quasiparticle energy relaxation.

In an ordinary superconductor near $T_c$, the charge imbalance relaxation time scales with the superconductor's intrinsic inelastic scattering time $\tau_E$ for electrons  \cite{tinkham_introduction_2004}
\begin{equation}\label{eq:taurelation}
\tau_{Q^*} = \frac{4  k_B T_C}{\pi \Delta(T)} \times \tau_E.
\end{equation}
This diverges near $T_c$ at the gap closes, which in turn yields a maximum in the charge imbalance length just below the transition. For $T\ll T_c$, it has been argued that this expression still holds to within a good approximation due to simultaneous energy relaxation and conversion between electron and hole-like quasiparticles \cite{tinkham_theory_1972}. As $T\rightarrow0$, $\Delta(T)$ becomes a constant and the time saturates to $\tau_E$ multiplied by a constant factor of order unity. Thus the charge imbalance time is expected to be determined solely by $\tau_E$ at millikelvin temperatures.  In earlier work on charge imbalance that was carried out near $T_c$, $\tau_E$ was found to be dominated by electron-phonon scattering\cite{clarke_experimental_1972,dynes_quasiparticle_1973}.  At the millikelvin temperatures relevant here, where electron-phonon scattering is greatly reduced, electron-electron scattering should be dominant. To understand this further, the effective inelastic rates $\tau_E^{-1}$ for our control device were calculated from $\lambda_{Q^*}$ and are also shown in Fig. \ref{fig:LambdaQStar}. The magnitude of these rates is consistent with the Nyquist electron-electron interaction rate for a 1D wire \cite{altshuler_effects_1982}, which supports the validity of this assumption for our control (see End Matter for further discussion). 

For proximity bilayers, determining $\tau_{e}$ becomes theoretically complicated. A similar case of charge imbalance in a bilayer consisting of two superconductors S and S' with different $T_c$'s has been considered theoretically by Golubov and Houwman \cite{golubov_quasiparticle_1993}. Their work found only a weak logarithmic modification of $\tau_{Q^*}$ due to proximity suppression of $T_c$. However, they also noted that $\tau_E$ for quasiparticles in a bilayer will be modified if the intrinsic inelastic relaxation time in the two layers is different. Their calculation gives that the effective inelastic time $\tau_{Eeff}$ is determined by a parallel contribution of the intrinsic inelastic times $\tau_{ES}$ and $\tau_{ES'}$ in the two layers:
\begin{align}\label{eq:Golubov}
    \frac{1}{\tau_{Eeff}}= \frac{1.82}{d_S}\left[ \frac{L_{eff}}{\tau_{ES}}+\frac{L'_{eff}}{\tau_{ES'}}\right]\frac{A'}{A},
\end{align}
where $d_S$ is the thickness of the superconductor with higher $T_c$ and $L_{eff}$/$L'_{eff}$ are effective thicknesses and $A$/$A'$ the area of the $S$ and $S'$ layers respectively. For a clean interface, they argue that the effective quasiparticle relaxation time is almost exclusively determined by the smaller of the two intrinsic relaxation times of the layers. Importantly, the basis of this quasiclassical theory should also be valid in the limit that that $T_c$ of the S' layer vanishes, i.e., the case relevant here when S' is a normal metal.

We believe that this effect explains the short $\lambda_{Q^*}$ seen in the bilayer device, and that the effective inelastic time for the bilayer (and consequently $\tau_{Q^*}$) is determined by the intrinsic inelastic time $\tau_E$ of the thin Cu layer alone, which is likely highly disordered and hence has a short $\tau_E$. In this scenario, it should be possible to tune $\tau_{Q^*}$ using a extremely thin NS bilayers so as to engineer faster quasiparticle relaxation while the entire device remains robustly superconducting. This is distinct from quasiparticle trapping, and relies only on enhancing the inelastic relaxation rate in one part of the bilayer. As this yields an observable decrease in an detected charge imbalance, this may also be a promising avenue for reducing the density of the bulk non-equilibrium quasiparticles responsible for quasiparticle poisoning without substantially reducing $T_c$ or generating large normal-metal regions across a qubit. This mechanism may be especially useful in the construction of Josephson junctions within qubits, as these are typically constructed from Al and could readily be adjusted to include a normal metal layer in their design. While in this work Cu was chosen primarily for its compatibility with Al, we also note that Cu is relatively easy to deposit via electron-beam evaporation, which may overcome some technical challenges related to heating during deposition reported previously for the fabrication of Ti/Al quasiparticle traps on Josephson junctions \cite{serniak_nonequilibrium_2019}. It may also be possible to improve upon this effect by selection of materials with higher intrinsic scattering rates or by the inclusion of specific defects. Further studies using both dc-transport and superconducting resonators or qubit measurements in disordered metal/superconductor bilayers will be required to determine the exact benefits of such an approach. 

In conclusion, we have observed a significant modification of the quasiparticle conductance within a superconducting Al wire via addition of a fully proximitized Cu underlayer. In  both samples we find that the superconducting density of states seen via tunneling can be explained via a simple Dynes-type fit, albeit with some apparent broadening due to an inhomogeneous gap. Using a dual-junction conductance measurement we find the absence of significant charge imbalance in the bilayer device compared to the control device. This effect can not be explained by pure $T_c$ suppression due to the inverse proximity effect, but by quasiclassical predictions indicating a strong tunability of the quasiparticle relaxation rate in a proximity effect bilayer. Given these findings, we predict that, for a sufficiently diffusive normal metal, the proximity effect could be used to artificially enhance the relaxation rate of quasiparticles in a superconductor without significantly reducing $\Delta$. Such enhancement may permit engineering of superconductors with lower quasiparticle populations at \SI{}{mK} temperatures to help address quasiparticle poisoning in superconducting qubits. 

\textit{Acknowledgments---} This material is based upon work supported by the U.S. Department of Energy, Office of Science, National Quantum Information Science Research Centers, Superconducting Quantum Materials and Systems Center (SQMS) under contract no. DE-AC02-07CH11359.

\bibliography{AlCu}% Produces the bibliography via BibTeX.

\onecolumngrid

\newpage
\section{End Matter}
\begin{figure}[b]
    \centering
    \includegraphics[width=0.48\textwidth]{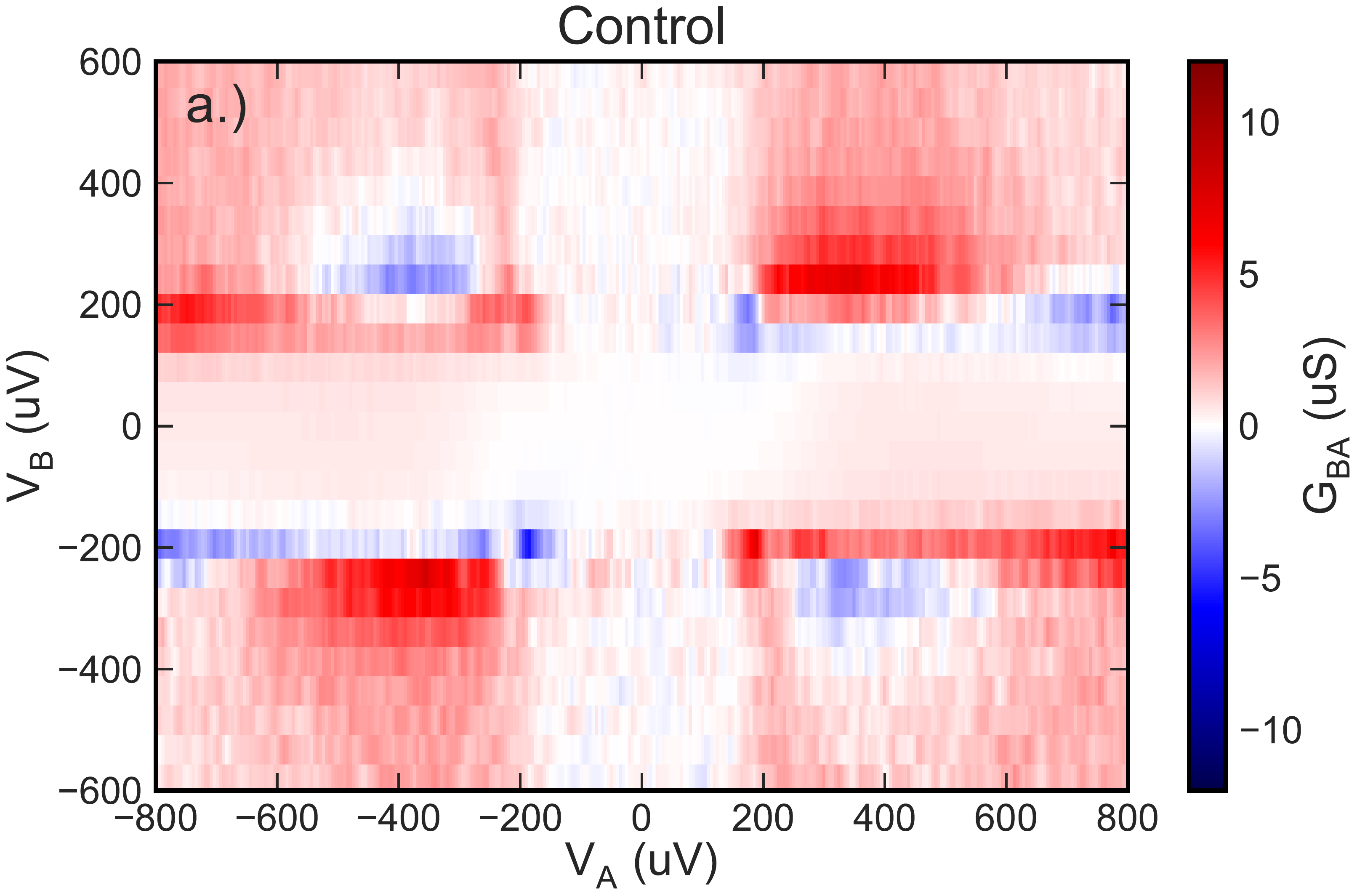}
    \includegraphics[width=0.48\textwidth]{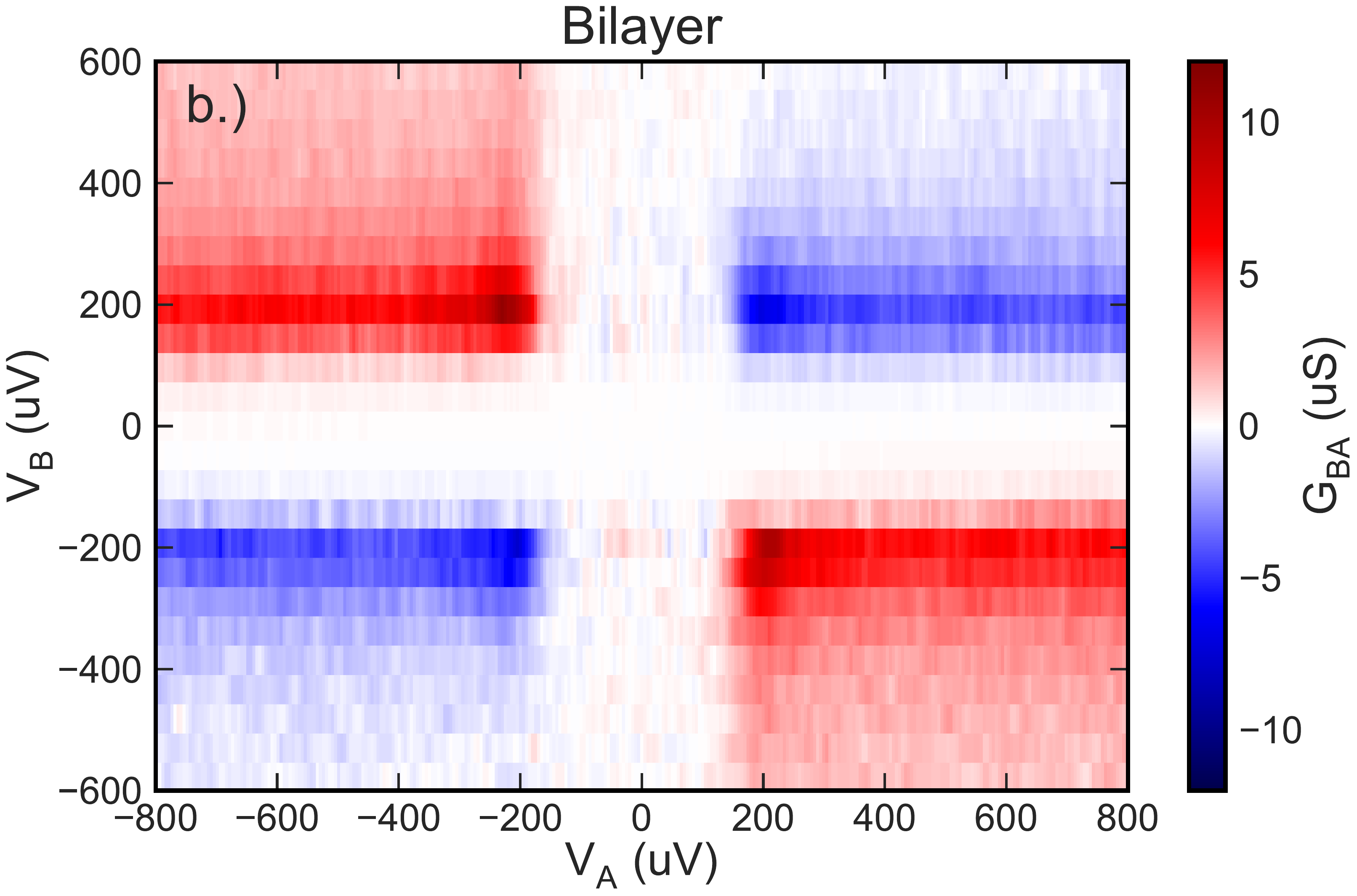}
    \caption{Dual bias conductance of a.) the control device and b.) the bialyer device, with junctions separated by $L = \SI{2}{\micro m}$, measured at \SI{20.7}{\milli\kelvin}.}
    \label{fig:heatmaps}
\end{figure}

\twocolumngrid
%\color{red}
\appendix*
\textit{Appendix A: Dual Bias Measurements---}
In addition to the charge imbalance measurements of the main text, measurements were performed on both the control and bilayer devices as a function of simultaneous dc biases of $G_{BA}(V_A,V_B)$. These data were taken using a circuit configuration identical to the one shown in Fig. \ref{fig:Schematics}, but with a voltage bias also applied to junction $B$ via the current pre-amplifier. By biasing the detector junction, we are no longer purely sensitive to the charge imbalance current, and instead also sample the non-equilibrium distribution of quasiparticles injected from junction A as they relax. These results are shown in Fig. \ref{fig:heatmaps}, and demonstrate several competing transport effects. Note that, the overall sign of the current $I_B$ is not determined by $G_{BA}$, and is generally dominated by the ordinary tunneling current when junction B is in the normal state. The sign convention for current follows that of the main text.

In the control device, $G_{BA}$ has a positive symmetric background as a function of both $V_A$ and $V_B$. This is consistent with charge imbalance, being enhanced somewhat in the normal state of the detector junction in the control device. In the bilayer device, the predominant behavior is anti-symmetric in both $V_A$ and $V_B$ at all bias ranges, with extrema near $|eV_A|$ or $|eV_B|\approx\Delta$. This is also seen at the detector junction gap edges when $|eV_B| \approx\Delta$. We do not expect this to be the result of quasiparticle charge transport, as the behavior matches well with reports by S. Kolenda et al., which they explain via the propagation of heat (generated in the normal-metal leads) across the device that in turn effects the temperature dependent tunneling across the detector \cite{kolenda_nonlocal_2013}.

Also present in the control device only are anti-symmetric ``lobes'' in $G_{BA}$, seen when $|eV_B|\approx\Delta$ and $|V_A|\approx\SI{400}{\micro V}$ with the opposite sign as predicted by S. Kolenda et al. Under these bias conditions, the detector junction samples the distribution of quasiparticles at or near the gap edge, at energies where they may not otherwise contribute to charge imbalance. As no corresponding conductance features are seen in the bilayer near the gap, we conclude that this may be a further indication that quasiparticles become fully relaxed and/or recombined before diffusing to the detector due to reduced relaxation time in the proximity effect bilayer.

\appendix*
\textit{Appendix B: Fabrication Details---}
\begin{table}[b]
\begin{tabular}{lccccc}
Device                      & Metal & Angle                           & $d$ (nm) & $P$ (Torr)       & Rate (\SI{}{Å\per\second}) \\ \hline
Control                     & Al    & $0^\circ$                       & 20.1   & \SI{4.0E-7}{}    & 1.1                        \\ \cline{2-6} 
                            & Cu    & $-35^\circ$                     & 80.2   & \SI{6.0E-7}{}  & 3.0                        \\ \cline{2-6} 
                            & Ti    & \multicolumn{1}{l}{$-35^\circ$} & 5.0    & \SI{2.8E-7}{}  & 0.5                        \\ \hline
\multicolumn{1}{c}{Bilayer} & Cu    & $-10^\circ$                      & 3.0    & \SI{1.0E-6}{}  & 1.2                        \\ \cline{2-6} 
\multicolumn{1}{c}{}        & Al    & $0^\circ$                       & 20.0   & \SI{6.9E-7}{}  & 1.05                       \\ \cline{2-6} 
\multicolumn{1}{c}{}        & Cu    & $-35^\circ$                     & 80.1   & \SI{1.0E-6}{} & 3.0                        \\ \cline{2-6} 
\multicolumn{1}{c}{}        & Ti    & $-35^\circ$                     & 5.0    & \SI{4.0E-7}{}  & 0.8                       
\end{tabular}
\caption{Deposition conditions during device fabrication. $d$ is the thickness determined during deposition, and $P$ is the recorded peak pressure during evaporation. }
\label{tab:Depo}
\end{table}
All metalization was performed using a multi-pocket e-beam evaporator with a tilting sample stage and a base pressure of \SI{1.8E-7}{Torr} or lower at ambient temperatures, with film thicknesses monitored via crystal monitor during deposition. The Cu, Ti and Al sources have purities of 4N, 4N5 and 5N respectively. To form the control device, deposition was performed of the Al at a $0^\circ$ incidence, followed immediately by dynamic oxidation in \SI{130}{mTorr} of flowing 99.994\% \ce{O2} gas for 15 minutes. Subsequently, \SI{80}{nm} of Cu was deposited upwards over the Al, followed by \SI{3}{nm} of Ti which was found to prevent oxidation of the Cu electrodes during liftoff and storage. To form the bilayer device, an identical procedure is followed except that \SI{3}{nm} of Cu was deposited before the Al at a small forward angle to prevent short circuiting to the normal metal leads. Care was taken to match deposition rates and oxidation conditions between the two devices, and the system was extensively pre-pumped to minimize active pressure during evaporation. These parameters are included in table \ref{tab:Depo}.

\color{black}
\appendix*
\textit{Appendix C: Scattering Rate Calculations---}To calculate the scattering rates obtained in Fig. \ref{fig:LambdaQStar} of the main text, we first determine the electronic mean free path $l_{mfp}$ of the control sample n the normal state from the sheet resistance $R_\square$ and established values for Al \cite{ashcroft2011solid}. With this we can calculate the normal electron diffusion coefficient by $D_N =  v_f l_{mpf}/3$. These quantities are reported in Tab. \ref{tab:Values}. 

\begin{table}[h]
\begin{tabular}{lccccc}
        & $R_\square$       & $w$                     & $d$                    & $l_{mfp}$               & $D_N$                                                                \\ \hline
Control & $\SI{6.21}{\ohm}$ & \SI{275}{\nano\meter} & \SI{20}{\nano\meter} & \SI{3.17}{\nano\meter}  & \SI[per-mode=reciprocal]{21.5}{\centi\meter\squared\per\second}  \\ \hline
Bilayer & $\SI{7.99}{\ohm}$ & \SI{300}{\nano\meter} & \SI{23}{\nano\meter} & $^\dagger$\SI{2.47}{\nano\meter} & $^\dagger$\SI[per-mode=reciprocal]{16.7}{\centi\meter\squared\per\second} \\ \hline
\end{tabular}
\caption{Measured and Drude model calculated values for both control and bilayer devices. $^\dagger$Approximate values for the bilayer calculated using the reported properties of Al alone. }
\label{tab:Values}
\end{table}

The diffusion coefficient for quasiparticles is energy dependent by Eq. \ref{eq:DiffusionCoeff} due to the energy dependence of their group velocity\cite{ullom_energy-dependent_1998}. To account for this, we approximate that the quasiparticle excitation energy $E_k= \sqrt{\epsilon_k^2 + \Delta^2}$ is $eV_A$. Taking this and the value of $\Delta$ obtained by the control fit performed in Fig. \ref{fig:ProxDOS}, we calculate the charge imbalance time to be 
\begin{equation}
    \tau_{Q^*}(V_A) \equiv \frac{\lambda_{Q^*}^2}{D_N} \frac{eV_A}{\sqrt{(eV_A)^2-\Delta^2}} ,
\end{equation}to which we apply Eq. \ref{eq:taurelation} using the measured  $\Delta$ and $T_c$ values reported in Figs. \ref{fig:ProxDOS} and \ref{fig:Tc} to estimate the inelastic rate $\tau_E^{-1}$. 

Interestingly, the inelastic scattering rates obtained by this calculation, while consistent in magnitude to other reports \cite{hubler_charge_2010}, are inconsistent with the rate of inelastic electron-electron scattering $\tau_{ee}^{-1}$ for Al at \SI{20}{\milli\kelvin}. Considering the predicted inelastic electron-electron scattering rate for a 1D wire \cite{PismaZhETF.33.515}
\begin{equation}
   \frac{1}{\tau_{ee}}  = \frac{1}{\pi^3} \frac{1}{\sqrt{2}} \frac{R_\square}{\hbar/e^2} \frac{D}{w L_T} , 
\end{equation}
written in terms of the Thouless length $L_T = \sqrt{\hbar D/k_B T}$, we find a rate $\tau_{ee}^{-1}$ for our control device of only \SI[per-mode=reciprocal]{3.0E5}{\per\second}. Thus, inelastic electron-electron scattering is insufficient to explain the relaxation of quasiparticles in our control device by almost three orders of magnitude.

We can instead consider the Nyquist electron-electron scattering rate \cite{altshuler_effects_1982} (which is not entirely inelastic):
\begin{equation}
    \frac{1}{\tau_N} = \bigg(\frac{1}{\sqrt{2}} \frac{R_\square}{\hbar/e^2}   \frac{\sqrt{D^3}}{ w L_T^2} \bigg)^{2/3}.
\end{equation}
Using the same values, we find a rate of \SI[per-mode=reciprocal]{3.4E7}{\per\second}. This is within a factor of 2-3 of the measured $\tau_E^{-1}$. This is curious, as the Nyquist rate is more closely related to de-phasing of electrons by small energy transfers \cite{wind_one-dimensional_1986} than to the energy relaxation mechanisms originally assumed by Tinkham \cite{tinkham_theory_1972} and similar works. A detailed study of the full temperature dependence of $\tau_{Q^*}$ would be necessary to understand these relationships further.

In light of the theory of Golubov and Houwman \cite{golubov_quasiparticle_1993}, predicting that the bilayer rate may be determined primarily by the fastest scattering rate of the two materials, we attempted to estimate the possible Nyquist rate of the Cu layer in our device. As the $R_\square$ value could not be measured independently, we instead deposited \SI{1}{\centi\meter\squared} Cu films of both \SI{3}{\nano\meter} and \SI{10}{\nano\meter} thicknesses at a rate of \SI{1.2}{Å\per\second} with active pressures during deposition of $\approx\SI{8E-7}{Torr}$ to closely match the films produced in the bilayer device. No capping layer was applied to the Cu. Tests of the \SI{3}{\nano\meter} Cu film indicated an immeasurably high sheet resistance, suggesting that the Cu film in our bilayer device may be below the percolation limit. Conversely, the \SI{10}{\nano\meter} Cu film had a sheet resistance of only $\SI{2.0}{\ohm}$, indicating the deposited film to be of fair quality and fully metallic. Both measurements were performed in air at ambient conditions immediately after removal from the deposition chamber to minimize oxidation of the Cu films. Extrapolating from this result, the \SI{3}{\nano\meter} Cu in our bilayer would have $R_\square\approx\SI{6.7}{\ohm}$, which corresponds to a diffusion constant of $\SI[per-mode=reciprocal]{170}{\centi\meter\squared\per\second}$. This yields a Nyquist rate of \SI[per-mode=reciprocal]{1.3E8}{\per\second}, an enhancement over the measured inelastic rate. However, this is likely to be a significant underestimation if the actual \SI{3}{\nano\meter} Cu film is below the percolation threshold, whereupon Coulomb blockade effects should give rise to a large enhancement of inelastic scattering in the bilayer which warrants further study.

% % \section{Conclusions}
\end{document}